\documentstyle[referee]{l-aa}
\begin{document}

\thesaurus{12(02.16.1;09.07.1;10.06.1)}

\title{Diffuse Gas Condensation Induced by Variations of the Ionizing Flux}

\author{Antonio Parravano\inst{1} and Catherine Pech\inst{2}}

\institute{
Centro de Astrof\'{\i}sica Te\'orica, Facultad de Ciencias,
Universidad de Los Andes, \\ A. Postal 26 La Hechicera, M\'erida 5251,
Venezuela. \\
\and Institut National des Sciences Appliqu\'ees de Toulouse, France.}

\offprints{A. Parravano}

\date{Received ???; accepted ???}

\maketitle

% \maintitlerunninghead{Induced condensation of diffuse gas}
% \authorrunninghead{A. Parravano et al.}
\markboth{Parravano \& Pech: Induced condensation of diffuse gas}{ ?? }

\begin{abstract}

The variation of an ionizing flux as a mechanism to stimulate the
condensation of a diffuse gas is considered. To illustrate this effect,
two situations are examined: one on the context of pregalactic conditions,
and the other on the context of the actual interstellar medium.
We focus our attention on flash-like variations; that is,
during a ``short'' period of time the ionizing flux is enhanced in
comparison to the pre- and post-flash values.
In both cases the cause of the induced phase change is the same:
the enhancement of the cooling rate by the increase in the electron density
caused by the momentary increase of ionizing flux.
After the passing of the flash, the cooling rate remains
enhanced due to the ``inertia of the ionization''. 
In the first case
(metal free gas) the cooling rate is enhanced due to the fact that
the increase of the electron density makes possible the gas phase
formation of $H_2$ by the creation of the intermediaries $H^-$ and
$H^+_2$. We show that after the passing of the photo-ionizing flash
a cloud near thermo-chemical equilibrium
at $\sim 8000\,\mbox{\rm K}$ may be induced to increase its $H_2$ content by many orders 
of magnitude, causing a rapid decrease of its temperature to values
as low as $100\,\mbox{\rm K}$.
In the second case (solar abundances gas) the dominant cooling mechanism
of the warm neutral gas (the excitation of heavy ions by
electron impacts) is proportional to the electron density. 
We show that, for the expected states of the warm interstellar gas, 
ionizing flashes may induce the phase transition from the warm to the cool
phase. 
The results indicate that the mechanism of induced condensation studied here
might play a relevant role in the gas evolution of the diffuse gas in both, 
the pregalactic and the actual interstellar medium conditions.
\end{abstract}

\section{Introduction}

Thermal condensation of diffuse gas is a commonly invoked route
for the formation of dense astrophysical structures.
This thermal condensation may occur spontaneously or
may be induced by variations of the external conditions. In the
first situation (spontaneous condensation), the initially diffuse
gas is in a state out of equilibrium in which the cooling dominates over
the heating.
Eventually, the gas reachs a thermo-chemical equilibrium in a dense and
cool state. In the second situation (induced condensation), in
absence of variation of external conditions the gas would reach (or be in) a
state of thermo-chemical equilibrium in a diffuse and hot state,
but in presence of appropriate variations of the external conditions, the gas
evolves toward a cooler and denser state.
Induced condensation of diffuse gas must play an important role
in the large scale evolution of the medium; specially if the
stimulating sources are able to induced condensation far away from them.
The synchronization and the large scale patterns of star formation may
be governed by this kind of stimulation.
In fact, most of the models and numerical simulations of the star formation 
in disk galaxies include as a main process the so-called self-propagating star 
formation 
(Gerola \& Seiden 1978; 
Seiden \& Gerola 1982; 
Shore 1981,
1983;
Palou\^{s} et al. 1990;
Comins \& Shore 1990; 
Cammerer \& Shchekinov 1994).
This kind of systems are part of the wider class of reaction-diffusion systems
(Kapral 1993, and references therein). In particular, the so-called exitable media
are appropriate to capture the main features of the pattern formation in disk
galaxies
(Smolin 1996).

Among the external conditions whose variations would induce phase
transitions are the external pressure and the ionizing flux.
We focus our attention on the effect of the variation of the ionizing flux,
keeping in mind that induced condensation by pressure variations may play a 
relevant role
(Roberts 1969;
Shapiro \& Kang 1987).
In particular, we focus our attention on flash-like variations; that is,
during a "short" period of time, the ionizing flux is enhanced in
comparison to the pre- and post-flash values.
To illustrate this mechanism of induced condensation,
two examples are given here: one on the context of pregalactic conditions
(in a free metal cloud), and the other on the context of the actual
interstellar medium (in a gas with solar abundances).
In both cases the cause of the induced phase change is the same:
the increase in the electron density due to the momentary increase of
ionizing flux enhances the cooling rate.
After the passing of the flash, the cooling rate remains
enhanced due to the inertia of the ionization (i.e. the characteristic
recombination time is much larger than the cooling one). 
\\
Many scenarios in which a flash of radiation affects the evolution of
gas clouds can be imagined. For the metal free gas case, this kind of
induced condensation might be relevant at the epoch of galaxy formation.
In particular, the ionizing flash effect might be a key step in the
sequence of events that have conduced to the formation of globular
clusters. It has been stated by Cox (1985) that ``such a rapid
fragmentation of the halo almost certainly requires inducement by
energy leaving the (primitive) disk''.
For the case of a gas with solar abundances, the evoked
mechanisms of induction are diverse but generally associated to compression
and convergent mass flows
(see review by Elmegreen 1992). 
However, in addition to these mechanisms, the induced condensation by 
ionizing flashes appears as an effect to be considered.

In Sec. 2 the basic equations to follow the thermal and chemical evolution
of a gas subject to variations of the gas pressure and the ionizing flux
are given.
In Sec. 3 the evolution of a free metal gas cloud subject to a flash of
ionizing and dissociating radiation is analyzed, whereas,
in Sec. 4, the evolution of a solar abundance gas subject to a flash
of cosmic ray flux is considered.
Finally, the conclusions are given in Sec. 5.

\section{Basic equations}

The thermal and chemical evolution of a gas subjected to variable external 
conditions are calculated by
solving the system of kinetic equations and the equation of energy
conservation. The system of kinetic equations can be written in the general form
\begin{equation}
\begin{array}{lll}
\frac{dX_i}{dt} & = & n\left(\sum_{j,k}\alpha_{jk}X_jX_k\,-\,
X_i \sum_{l}\alpha_{il}X_l\right) \\
 & & \\
 & & +\,\sum_{m}\alpha_mX_m\,-\,\alpha_i X_i\, ,
\end{array}
\end{equation}
where $n$ is the total number density of hydrogen nuclei and $X_i$ is the
relative number density of the $i$-th species. 
In Eq. (1), $\alpha_{jk}$ and $\alpha_{il}$ are, respectively, the formation and
destruction rates of the $i$-th species due to double collisions, whereas,
$\alpha_m$ and $\alpha_i$ are, respectively, the production and destruction rates
of the $i$-th species due to the interaction of particles with radiation.

The equation of energy conservation equation can
be expressed as
\begin{equation}
\frac{dU}{dt}+P \frac{dV}{dt} + {\cal L} = 0\, ,
\end{equation}
where U is the internal energy by unit mass, $\cal{L}$  is the net cooling rate 
by unit mass ($\cal{L}$ $=(\Lambda - \Gamma)/\rho$), $P$ is the pressure,
$\rho$ is the mass density, and  $V$ is the specific volume.
Assuming that $P=\rho R T /\mu$ and $U=3 R T/ 2 {\mu}$, with R the gas constant, 
and $\mu$ the molar mass, eq. (2) can be written as
\begin{equation}
\frac{dT}{dt}=\frac{T}{\mu}\,\frac{d\mu}{dt}+ \frac{2}{5}\,\frac{T}{P}\,\frac{dP}{dt}
\,-\frac{2}{5}\,\frac{\mu}{R}\,\cal{L}\, .
\end{equation}

Two extreme situations that can be used to confine intermediate situations
are the constant density and constant pressure approximations.
In these approximations, the energy conservation equation can
be expressed as
\begin{equation}
\frac{dT}{dt}=\frac{T}{\mu}\,\frac{d\mu}{dt}-\lambda\,\frac{\mu}{R}\,\cal{L}\, ,
\end{equation}
where, $\lambda=2/3$ or $2/5$ in the constant density or in the constant pressure 
approximations, respectively.

Variations of the external ionizing sources flux produces direct 
variation on the rates 
$\alpha_m$ and $\alpha_i$, and on the heating rate $\Gamma$, whereas, 
variations of the presure affect density and temperature.
Notice that variations of $n$ or $\alpha_i$ provoke a variation of the 
chemical state of the gas, and therefore produce a variation of the
cooling rate $\Lambda$. In the following, we solve the basic equations
(1) and (2) with the appropriate assumptions in order to model: a) the
evolution of a metal free cloud subjected to a variation of the ionizing
and dissociating flux (sec. 3), and b) the evolution of a solar abundance gas
subjected to variations of the primary ionization rate due to cosmic
rays (sec. 4).

\section{Photo-ionizing radiation flash as trigger of efficient
$H_2$ cooling in free metal gas clouds}

In the context of formation of galaxies and globular clusters, the
non-equilibrium formation of $H_2$ has been identified as a key process for
achieving a rapid cooling bellow $10^4 \mbox{\rm K}$ 
(Palla \& Stahler 1983;
Izotov \& Kolesnik 1984;
Shapiro \& Kang 1987; 
Palla \& Zinnecker 1987; 
Kang et al. 1990; 
Anninos et al. 1996; 
Padoan et al. 1996;
Tegmark et al. 1996).
On the other hand, the ionization and dissociation of primordial gas by UV 
background radiation have been taken into account as an important parameter in
many studies of the thermo-chemical evolution of pregalactic and intergalactic
structures 
(Kang et al. 1990; 
Donahue \& Shull 1991; 
Ferrara \& Giallongo 1996;
Haardt \& Madau 1996;
Navarro \& Steinmetz 1996;
Mucket \& Kates 1997). 
In general, the UV background radiation acts as an inhibitor
of $H_2$ formation. However, there are exceptions to this rule.
Recently 
Haiman et al. (1996) 
have shown that UV 
background radiation can enhance the formation of $H_2$ in primordial gas 
at high densities ($\geq 1 \mbox{\rm cm}^{-3}$) and low temperatures ($\leq 10^4 \mbox{\rm K}$); 
but for densities lower than the above value, the effect of a {\it constant} 
UV background radiation is to inhibit the $H_2$ formation.
As it will be shown in the present section, rapid variations of the UV 
background radiation can also enhance the non-equilibrium formation of $H_2$.
In particular, we focus our attention on flash like variations capable of
heating and increasing the ionization fraction of an initially warm-neutral 
gas cloud near thermo-chemical equilibrium.
After the passing of the radiation pulse, the enhanced ion fraction makes
the gas phase formation of $H_2$ molecules possible by the creation of the
intermediaries $H^-$ and $H^+_2$.
The presence of small quantities
of $H_2$ molecules then makes possible further radiative cooling to
temperatures as low as $\sim 10^2\,\mbox{\rm K}$. The rapid cooling ($\sim 0.2$
free-fall times) abruptly reduces the Jeans mass by a factor $\sim 10^4$, 
permitting the fragmentation of clouds initially marginally stable.
Izotov (1989) has considered the homogeneous contraction approach of a
gravitational unstable cloud, but here we are interested in delimiting the
necessary conditions that provoke the rapid cooling of the cloud without
invoking the gravitational collapse.
UV radiation pulses have similar effects than shock waves because 
the post-shock flow also recombines out of equilibrium.

The spectrum, amplitude and duration of the radiation
pulse are free parameters in our model.
However, it should be noticed that,
in order to induce efficient $H_2$ cooling in the cloud, the detailed form 
of the flash is not important as long as that, during the
flash, the ionization of the cloud increases appreciably
and after this, the ionizing flux decreases to background values
in a short time compared to the recombination time.

The initial conditions of the cloud and the characteristics of the 
hypothetic ionizing pulse depend on the chosen scenery.
For example in the Fall and Rees (1985) scenery, 
during the proto-galactic collapse, if the gas is assumed to be lumpy,
the overdense regions will cool more rapidly than the underdense regions
producing a two-phase medium. But the developing
rate of this overdense regions depends on the initial conditions
(i.e. the denser regions develop faster; Murray and Lin 1990) and on
their interaction with the system.
Thus, a dependence of the density contrast and of the mean cloud masses on the
galactocentric distance are expected.
The assumption of an ionizing flash produced in the galactic nucleus
implicitly assumes a radial increment of the delay in the evolution
of the clouds relative to the center.
The dilution and attenuation of the ionizing and dissociating radiation
produced in the flash also introduce a radial dependence.
In fact, the possibility that proto-galactic structures are exposed to
UV radiation emitted by massive young stars or an active galactic nucleus
has been considered previously (Kang et al. 1990).
Therefore, it would be useful to know the dependence of the cloud evolution on
the flash characteristics, and on the cloud initial state
(i.e cloud mass ($M_c$), temperature ($T$), number density ($n$), and the
relative number density of the species ($X_i$).

To follow the thermal and chemical evolution of a metal-free cloud in presence
of a variable radiation flux, we consider an idealized uniform cloud.
The gas model adopted for this application considers 
the following 9 species: 
$H$, $H^+$, $e$, $H^-$, $H_2^+$, $H_2$, $H_e$, $H_e^{+}$, and $H_e^{++}$.
Assuming a mass fraction $y$ ($=0.3$) of Helium relative to Hydrogen, then
$\mu=(1+y)/(1+X_e-X_{H_2}-X_{H_2}^+ +y/3.97)$, where $X_i$ is the ratio of the
number density of species $i$ to the total number density of Hydrogen nuclei $n$.
The processes of formation and destruction of these nine species are assumed to be
the 24 reactions in Table I of Rosenzweig et al. 1994.
The adopted cooling rates, include:
1) the collisional ionization of $H$ by electron impact; 
2) the free-free transitions of $H^{+}$ (Izotov (1989) and references
therein); 
3) the collisional ionization by electron impact of $H_e$ and $H_e^{+}$;
4) the free-free transitions of $H_e^{+}$ and $H_e^{++}$;
5) the total dielectronic cooling rate of $H_e^{+}$ (Shapiro \& Kang 1987);
6) the $H_{2}$ cooling rates due to rotational and vibrational transitions
excited by ($H - H_2$) and ($H_{2} - H_{2}$) collisions, calculated according
to Lepp and Shull (1983). These $H_{2}$ cooling rates must be multiplied
by a factor
$L(\tau_{vib})$ and $L(\tau_{rot})$
to take into account the escape probability of the vibrational and rotational
photons, respectively. 
Finally, 7) the excitation
of the $H_{2}$ vibrational levels ($v\leq4$) by low energy electron-impact
was taken into account (Klonover \& Kaldor, 1979).\\

The external radiation flux affects the photoionization of the Hydrogen 
and Helium, 
and the $H_2$ photo-dissociation (and its associated heating rates; 
see Rosenzweig et al. 1994). The external photon flux is assumed to have the 
quasar-like distribution (Shapiro and Kang 1987)
\begin{equation}
{\cal {N}}(\nu,t)=4.6\times 10^{-11}\,\epsilon(t)\,
\left(\frac{\nu}{\nu_H}\right)^{-1.7}\:\mbox{\rm cm}^{-2}\,\mbox{\rm Hz}^{-1}\,\mbox{\rm s}^{-1}\, ,
\end{equation}
where $\nu_H$ is the H Lyman-edge frequency. The function $\epsilon (t)$ is 
introduced to represent the changes of the flux level outside the cloud, and
is assumed to behave as $\epsilon_{ion}(t)$ in the Lyman continuum, and as
$\epsilon_{dis}(t)$ for the photo-dissociating photons.
Since a homogeneous cloud is assumed in this application, the optical depth 
at its center, for the ionizing radiation, can be written as: 
\begin{equation}
\tau _{\nu} = 6.6 \times 10^{18} n (\sigma _{\nu}^{H} X_{H} 
+ \sigma _{\nu}^{H_e} X_{H_e} + \sigma _{\nu}^{H_e^+} X_{H_e^+}) 
(\frac{M}{n})^{1/3}
\end{equation}
where $M$ is the mass of the cloud in units of $\mbox{\rm M}_{\odot}$ and 
$\sigma _{\nu}^i$ are the absorption cross sections of species $i$ at 
frequency $\nu$.
As a first approximation, the attenuation of the external radiation field 
${\cal {N}}(\nu,t)$ by the factor $\exp ^{-\tau _{\nu}}$ is adopted 
to schematize the opacity effects. For the dissociating radiation field
a similar approximation is adopted (Rosenzweig et al. 1994).
Obviously, by using this first approximation, the effect of the ionizing radiation
is minimized, and the study of the internal structure
of the cloud is not possible. 

For the time dependence of the flux level $\epsilon$ we assume a 
flash like variation.
More precisely, during a first period of integration the external radiation flux
remains at the assumed background value 
(i.e. $\epsilon_{ion}=\epsilon_{dis}=\epsilon_{0}$).
The duration of this initial period of integration is large enough 
(i.e. $\sim 10^8$ yr) to ensure that the cloud is near its thermo-chemical 
equilibrium state. 
After this first period of stabilization, it 
is assumed that both, $\epsilon_{ion}$ and $\epsilon_{dis}$,
increase linearly to the value $\epsilon_{1}$ in a time $\tau_{incr}$, 
and remains at this value during a time $\tau_{flash}$.
Then, the flux level is assumed to decrease 
exponentially with a characteristic time $\tau_{decr}$ to the background value 
$\epsilon_{0}$ for the ionizing flux, and to $\epsilon_2$ for the dissociating flux. 
This kind of flux variations are expected in the case
when the radiation pulse is assumed
to be produced by an intense but short event of stellar formation
or by the radiation coming from a front shock. In a
few million years after the end of the stellar formation process, the 
ionizing flux is expected to decrease to background values. 
On the other hand, due to the contribution of intermediate and low mass stars
to the dissociating flux, it is expected that its level remain higher
than the background value ($\epsilon_2 > \epsilon_0$) long after the death of
massive stars. 

For a given initial state of the cloud, there is a critical background
flux level $\epsilon_{cri}$, bellow which the cloud spontaneously cools because
the self-shielding allows the formation of $H_2$ at relatively high concentrations.
This critical background flux level is mainly sensible to the initial
ion fraction due to its strong influence on the $H_2$ formation rate.
In order to describe this initial state dependence, it is useful to look
at the thermo-chemical equilibrium (TCE) curve
(i.e. ${\cal{L}} = 0$, and $\frac{dX_i}{dt} = 0$).
Figure (1.a) shows the TCE curves for two different values of the 
background flux level (lower curve $\epsilon_0 = 1$, and
upper curve $\epsilon_0 = 10$)
when the mass of the cloud is $10^7\, \mbox{\rm M}_{\odot}$.
\setcounter{figure}{0}
\renewcommand{\thefigure}{1.a}
\begin{figure}[htb]
\vspace{7.8cm}
\includegraphics{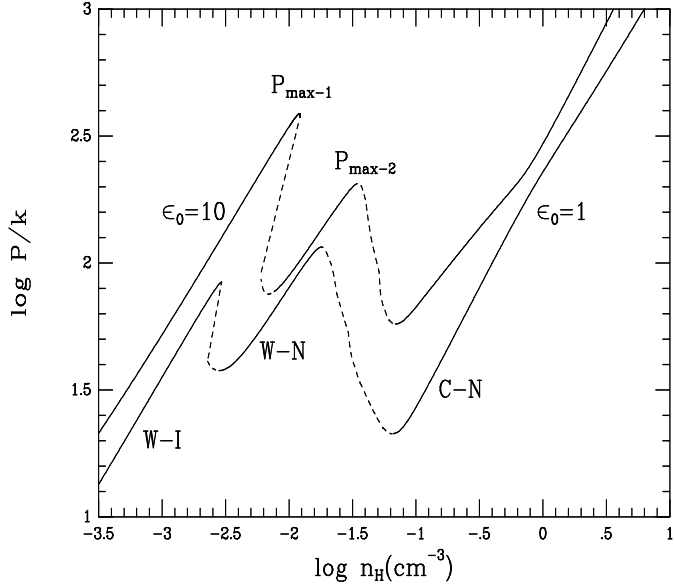}
\caption[]{The thermo-chemical equilibrium curves for a $10^7\, \mbox{\rm M}_{\odot}$
cloud exposed to the two labeled values of the background flux level (see text).
}
\end{figure}
\setcounter{figure}{0}
\renewcommand{\thefigure}{1.b}
\begin{figure}[htb]
\vspace{7.8cm}
\includegraphics{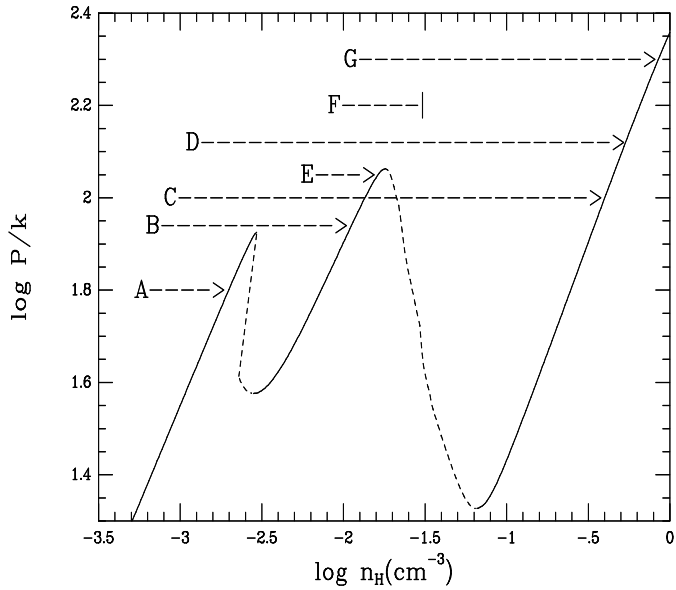}
\caption[]{The thermo-chemical equilibrium curve and the typical isobaric evolution tracks
for initially ionized states (arrows A, B, C and D) and for initially neutral states
(arrows E, F and G).
}
\end{figure}
Note the three phase structure characterized by the presence of three stable
branches (solid lines) denoted in Fig. (1.a) as
W-I (i.e. warm-Ionized; $T \ga  13000 \mbox{\rm K}$),
W-N (i.e. warm-Neutral; $8000 \mbox{\rm K} \la  T  \la  6000 \mbox{\rm K}$), and
C-N (i.e. cool-Neutral; $T \la 300 \mbox{\rm K}$).
Note also the qualitative difference of the TCE curve
for low ($\epsilon_0 = 1$) and high ($\epsilon_0 = 10$) background flux level. That is,
for $\epsilon_0 = 10$ the left-hand maximum (denoted as $P_{max-1}$) is over the
right-hand maximum (denoted as $P_{max-2}$), whereas, 
$P_{max-1}<P_{max-2}$ for $\epsilon_0 = 1$.
This difference is important because if the cloud
is initially in TCE in the W-I branch, and
the pressure is progressively increased from $P< P_{max-1}$ to $P > P_{max-1}$
a transition to the W-N phase occurs in the case $\epsilon_0 = 1$, but in the
case $\epsilon_0 = 10$ the C-N phase is reached.
For $\epsilon_0 = 1$, Fig. (1.b) shows the typical isobaric evolution tracks for initially
ionized states (arrows A, B, C and D) and for initially neutral states
(arrows E, F and G). For cases A, B, and D,  as expected the cloud evolves toward
the stable branch W-I, W-N, and C-N, respectively. However, in case C, even when
$P< P_{max-2}$ the track reaches the C-N branch transversing the W-N branch due to
the inertia of the ionization fraction. More precisely, the excess of the
ionization fraction shifts the maximum pressure $P_{max-2}$
of the corresponding thermal
equilibrium curve to a value bellow the pressure corresponding to case C.
On the other hand, for the initially neutral states the evolution is 
substantially different. 
In case E the track stops in the W-N branch; note the difference
with case C. In case F the cooling rate is so small that in practice the cloud
reaches a quasi-stationary state similar to that in the W-N branch. The
small cooling efficiency is due to the very small $H_2$ formation rate for the
low ion fraction in the quasi-stationary W-N state. Finally, for high enough
gas pressure as in case G, the cloud reaches the C-N branch.
The isobaric evolution tracks for a high background flux level (i.e. $\epsilon_0 = 10$)
are similar to those in Fig (1.b) except that the W-N branch can not be reached
from an ionized initial state as in case B in Fig. (1.b).
It is interesting to note that when the cloud is in the quasi-stationary state
schematized by track F, a large enough  ionizing flash is able to increase 
$H_2$ formation rate
and to stimulate the rapid condensation to reach the C-N branch.

\setcounter{figure}{0}
\renewcommand{\thefigure}{2.a}
\begin{figure}[htb]
\vspace{7.8cm}
\includegraphics{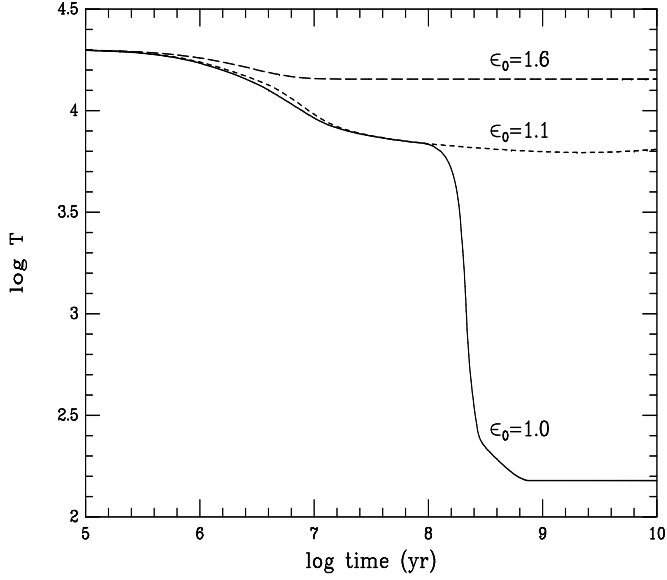}
\caption[]{The time dependence of the temperature for a $10^7\, \mbox{\rm M}_{\odot}$
cloud exposed to the three labeled values of the background flux level
and an initial out of equilibrium warm-ionized state (see text).
}
\end{figure}
\setcounter{figure}{0}
\renewcommand{\thefigure}{2.b}
\begin{figure}
\vspace{7.8cm}
\includegraphics{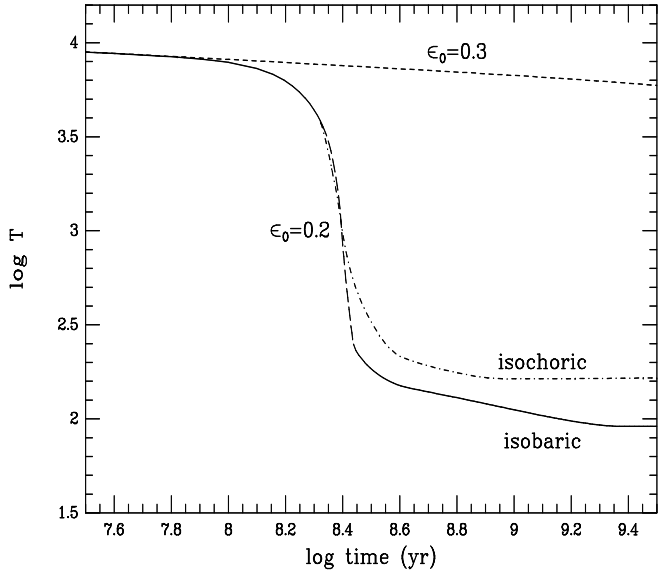}
\caption[]{The time dependence of the temperature for a $10^7\, \mbox{\rm M}_{\odot}$
cloud exposed to the two labeled values of the background flux level
and an initial out of equilibrium warm-neutral state (see text).
}
\end{figure}

The analysis of the various evolution tracks in Fig. (1.b) have been made for
a fixed background flux level. Taken into account that an increase of the
background flux level shift the TCE curve upward, a similar analysis
can be made for a fixed pressure but varying the background flux level.
Figure (2.a) shows, in the constant pressure approximation, the time dependence
of the gas temperature for a
$10^7\, \mbox{\rm M}_{\odot}$ cloud subject to the three labeled values of the
background flux level $\epsilon_0$. The initial condition is the
same in the three cases and corresponds to an out of equilibrium warm-ionized 
state at a gas pressure $log(P/k)=2.046$, and
$n=2.5 \times 10^{-3} \mbox{\rm cm}^{-3}$,
$T=2 \times 10^4 \mbox{\rm K}$,
$X_{H}=8.8 \times 10^{-4}$,
$X_{H_2}=2.3 \times 10^{-13}$,
$X_{H^+}=0.999$,
$X_{e}=1.147$,
$X_{H^-}=4.6 \times 10^{-11}$,
$X_{H_2^+}=1.2 \times 10^{-11}$,
$X_{He}=2.57 \times 10^{-6}$,
$X_{He^{+}}=3.11 \times 10^{-3}$, and
$X_{He^{++}}=7.24 \times 10^{-2}$.
The temperature evolution in the three plotted cases differs because for
$\epsilon_0=1.6$ the system stabilizes in the W-I branch
(i.e. as in track A of Fig. (1.b)),
for $\epsilon_0=1.1$ the system stabilizes in the W-N branch
(i.e. as in track B of Fig. (1.b)), and
for $\epsilon_0=1.0$ the system stabilizes in the C-N branch
(i.e. as in track C of Fig. (1.b)).
On the other hand, Fig. (2.b) shows the temperature evolution for the same
pressure as in Fig. (2.a), but for an initially neutral warm state, (i.e.
$n=1.0 \times 10^{-2} \mbox{\rm cm}^{-3}$,
$T=1.03 \times 10^4 \mbox{\rm K}$,
$X_{H}=0.9992$,
$X_{H_2}=5.801 \times 10^{-7}$,
$X_{H^+}=7.978 \times 10^{-4}$,
$X_{e}=2.631 \times 10^{-3}$,
$X_{H^-}=3.914 \times 10^{-8}$,
$X_{H_2^+}=4.813 \times 10^{-9}$,
$X_{He}=7.374 \times 10^{-2}$,
$X_{He^{+}}=1.819 \times 10^{-3}$, and
$X_{He^{++}}=7.169 \times 10^{-6}$).
In this case, the evolution for $\epsilon_0=0.3$ corresponds to the
situation schematized by track F in Fig. (1.b); that is, the system
enters in a quasi-stationary state characterized by a very slow
decrease of the temperature. However, for $\epsilon_0=0.2$ the
system evolves rapidly to the C-N branch, a situation that
corresponds to that schematized by track G in Fig. (1.b).
The results in Figs. (2.a) and (2.b) illustrate how the
initial cloud state affects the critical value of $\epsilon_0$
bellow which the cloud evolves toward the C-N branch (i.e. the cloud
condenses).
In order to show the effect of the constant density approximation,
Fig. (2.b) also shows the temperature evolution for
$\epsilon_0=0.2$ when the constant pressure approximation is
switched to the constant density approximation at the time when
the rate of decrement of the cloud radius equals the sound speed.

\setcounter{figure}{0}
\renewcommand{\thefigure}{3}
\begin{figure}[htb]
\vspace{7.8cm}
\includegraphics{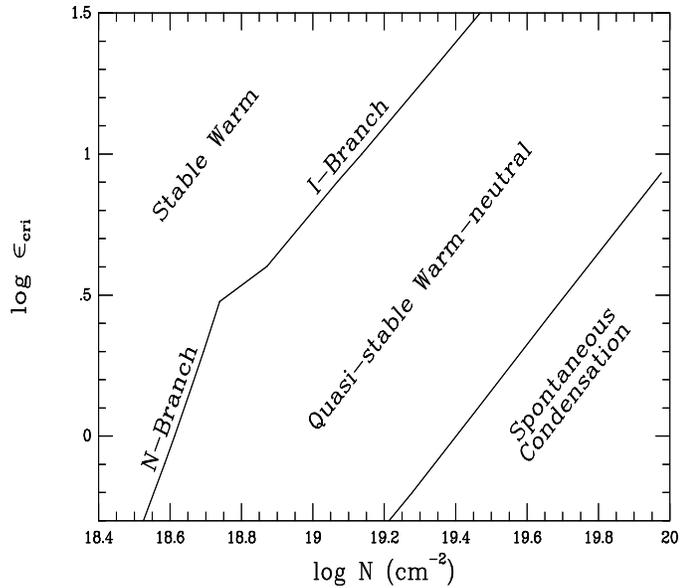}
\caption[]{The dependence of $\epsilon_{cri}$ on the initial cloud column
density (see text).
}
\end{figure}

For initially ionized states at pressure $P$, the critical value
$\epsilon_{cri}$ roughly corresponds to the value of $\epsilon_0$ that produces
a maximum in the TCE curve at pressure $P$.
For a high background flux level, the left-hand maximum (denoted as
$P_{max-1}$ in Fig. (1.a)) is the higher maximum and must be equaled to $P$.
For a low background flux level, the right-hand maximum must be equaled to $P$.
On the other hand, for initially neutral states, the critical value
$\epsilon_{cri}$ must be calculated by finding the value
of $\epsilon_0$ bellow which the evolution does not fall in the
quasi-stationary warm-neutral state.
The dependence of the critical value of $\epsilon_{cri}$ on the initial cloud
state is summarized in Fig. (3) for a $10^7\, \mbox{\rm M}_{\odot}$ cloud.
The results are showed as function of the initial cloud column
density ($N=n \times R_{cloud}$). The left-hand side curve corresponds to the
critical background flux level for initially ionized states. The upper
segment (labeled I-branch) corresponds to background flux levels for which
$P_{max-1}>P_{max-2}$, and therefore, the condition 
$P_{max-1}(\epsilon_{cri})=P$ is used.
The lower segment (labeled N-branch) corresponds to the case when the 
condition $P_{max-2}(\epsilon_{cri})=P$ is used because $P_{max-2}>P_{max-1}$ 
for these values of $\epsilon_{cri}$.
The initial column density is calculated assuming that initially
the cloud is at $2.0 \times 10^4 \mbox{\rm K}$ and at a pressure 
$P=P_{max}(\epsilon_{cri})$;
the initial concentrations are assumed to be the used in Fig. (2.a).
For values of ($N,\epsilon_0$) above the left-hand $\epsilon_{cri}(N)$ curve,
the initially ionized cloud reaches TCE in the stable warm phase, but below
this curve the cool-neutral branch is reached.
On the other hand, the right-hand side curve corresponds to the
critical background flux level for initially neutral states. In this case,
the initial column density is calculated assuming that initially
the cloud is at $1.2 \times 10^4 \mbox{\rm K}$ and the initial concentrations
are assumed to be the used in Fig. (2.b); the initial density is varied
in order to cover the plotted range of $\epsilon_{cri}$.
For values of ($N,\epsilon_0$) above the right-hand $\epsilon_{cri}(N)$ curve,
the initially neutral cloud attains a quasi-stationary warm-neutral state,
but below this curve the cool-Neutral branch is reached. 
It is to be noticed that as the initial electron concentration increase,
the right-hand $\epsilon_{cri}(N)$ curve approaches the left-hand curve.
In any case, there exist a set of
initial conditions for which the cloud reaches a quasi-stationary 
warm-neutral state. The point to be emphasized is that these 
quasi-stationary states  are susceptible to be
induced to condensate if the cloud is exposed to an intense enough 
ionizing flash.

\setcounter{figure}{0}
\renewcommand{\thefigure}{4a}
\begin{figure}[htb]
\vspace{7.8cm}
\includegraphics{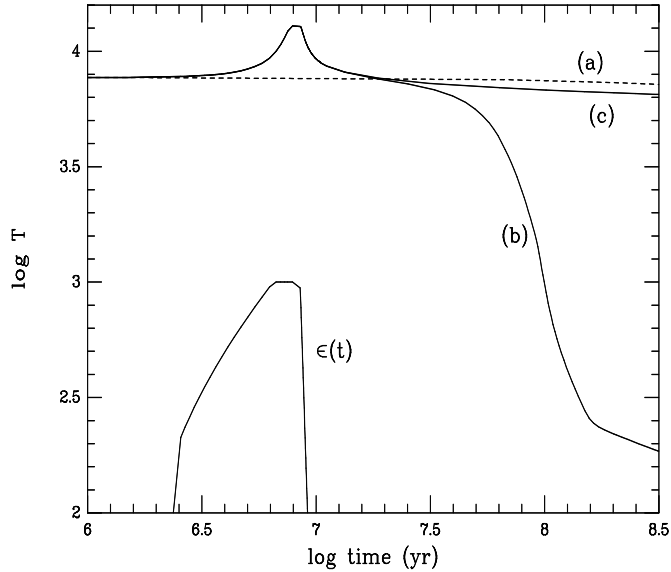}
\caption[]{The effect of the ionizing flash $\epsilon (t)$ on the
the time dependence of the temperature for a $10^7\, \mbox{\rm M}_{\odot}$ cloud.
The curve labeled (a) corresponds to the case
$\epsilon(t)=\epsilon _{0} > \epsilon_{cri}$.
The curves labeled (b) and (c) correspond to the evolution of
the cloud when it is subjected at $t=2\times 10^8$ yr to a flash (see text).
}
\end{figure}
\setcounter{figure}{0}
\renewcommand{\thefigure}{4b}
\begin{figure}[htb]
\vspace{7.8cm}
\includegraphics{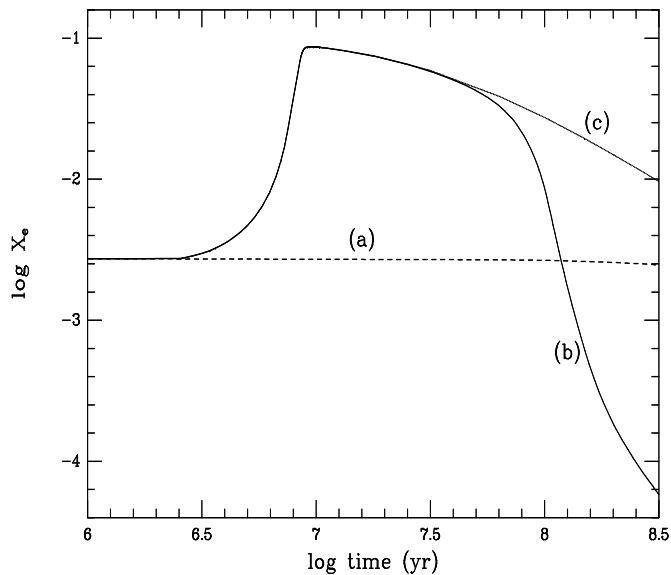}
\caption[]{The time dependence of  the relative number density of
electrons ($X_{e^-}$) for the initial and external conditions corresponding to
Fig. (4a).
}
\end{figure}
\setcounter{figure}{0}
\renewcommand{\thefigure}{4c}
\begin{figure}[htb]
\vspace{7.7cm}
\includegraphics{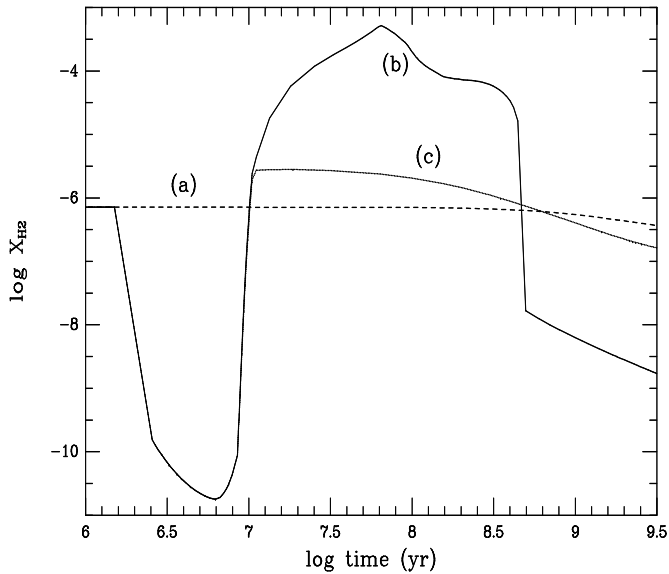}
\caption[]{The time dependence of the Hydrogen molecules ($X_{H_2}$),
for the initial and external conditions corresponding to Fig. (4a).
}
\end{figure}

In order to illustrate the effect of the ionizing flash on a cloud that
has fallen in the quasi-stationary warm-neutral state, Figs. (4a,b,c)
show respectively the time dependence
of the temperature, the relative number density of
electrons ($X_{e^-}$), and of Hydrogen molecules ($X_{H_2}$),
for a $10^7\, \mbox{\rm M}_{\odot}$ cloud.
The initial condition corresponds to an out of equilibrium warm-neutral state
at a gas pressure $log(P/k)=2.074$ (i.e. $n=1.1 \times 10^{-2} \mbox{\rm cm}^{-3}$,
$T=1.0 \times 10^4 \mbox{\rm K}$, and the initial concentrations used in Fig. (2.a)).
The curves labeled (a) in Figs. (4) are plotted for reference, and correspond
to the case when the cloud is subjet to a constant background flux level
of $\epsilon _{0}= 1.0$.
As expected for $\epsilon(t)=\epsilon _{0} > \epsilon_{cri}$,
the cloud evolves toward a quasi-equilibrium warm neutral state.
The curves labeled (b) and (c) in Figs. (4) correspond to the evolution of
the cloud when it is subjected at $t=2\times 10^8$ yr to a flash.
In case (b), the flash characteristics are:
$\tau_{incr}=5\times 10^6$ yr, $\tau_{flash}=2\times 10^6$ yr,
$\tau_{decr}=2\times 10^6$ yr,
$\epsilon_1= 10^3$, and $\epsilon_2=1.5 \times \epsilon _{0}$.
In case (c) the flash characteristics are the same as in case (b) but
with $\epsilon_2=2\times \epsilon _{0}$.
Notice in Figs. (4) that during the increase of the external radiation flux
from $\epsilon _{0}$ to $\epsilon_1$, the electron density and
the temperature increase, whereas the $H_2$ and $H^-$ densities decrease.
During the time when $\epsilon(t)=\epsilon _{1}$, the electron density and
the temperature continue to increase because the variation of $\epsilon(t)$
during the time $\tau_{incr}$ is rapid enough to leave the gas far from
equilibrium.
At the end of the lapse of decrement of $\epsilon(t)$, if $\tau_{decr}$ is
short enough, the gas has an excess of electrons and thermal energy compared
with the equilibrium values corresponding to $\epsilon _{ion}=\epsilon _{0}$
and $\epsilon _{diss}=\epsilon _{2}$.
After the lapse of UV flux decrement, recombination
continues, but at a lower rate than the cooling.
The excess of electrons at these relatively low temperatures results
in an enhancement of the $H_2$ rate formation.
Even when $\epsilon _{2} > \epsilon _{0}$, the abundance of $H_2$ may reach
a large enough value to produce considerable self-shielding.
If the cloud reaches a critical value ($\tau \sim 1$) for the optical depth
at dissociating
frequencies, the $H_2$ abundance grows very fast, allowing the cooling
of the cloud to temperatures of the order of $100 \mbox{\rm K}$.
This is the situation for case (b) in Figs. (4), where the flash induce rapid
cooling even when $\epsilon _{2} = 1.5 \times \epsilon _{0}$.
On the other hand, for case (c) a post flash dissociating level with
$\epsilon _{2} = 2 \times \epsilon _{0}$ is enough to inhibit the formation
of $H_2$, and then, the cloud remains warm.

A detailed study of the dependence of the cloud evolution on the phase
space of free parameters (cloud mass, initial conditions, and flash
characteristics) is out of the scope of this simple application.
However, such detailed study may reveal that the stimulating condensation
process studied here can be effective in a restricted region of the
free parameters space, and therefore, may act as a selective effect that
contributes to the formation of dense structures at certain scales.

\section{Ionizing pulse as trigger of warm to cool phase transition in a
gas with solar abundances}

Among the various processes that determine the change of state of the
ISM material, phase transitions are expected to play an important
role. In particular, warm to cool phase transition has been indicated 
as a channel to transform diffuse gas ($< 10^{-1} \mbox{\rm cm}^{-3}$) into denser
states ($>10^{1} \mbox{\rm cm}^{-3}$) 
(Field, Goldsmith \& Habing 1969;
Lepp et al. 1985;
Parravano 1987;
Lioure \& Chi\`{e}ze 1990;
Dickey \& Brinks 1993).
Moreover, it has been proposed (Parravano 1988, 1989) that the large scale 
star formation rate must be self-regulated because, on one hand the warm gas 
condensation is inhibited by high enough UV radiation (coming mainly from 
massive stars), and on the other hand, a gas supply from the diffuse phases 
is required to feed the large scale star formation process. 
In this way, the large scale star formation rate is limitated, and the warm 
gas tends to remains close the critical state for warm to cool phase transition.
The fact that large quantities of ISM warm gas remain close to this critical
state allows the trigger of condensation
by relatively small variations of the ambient conditions. 
More precisely, the mean ``distance'' of the warm gas state to the critical 
state for spontaneous condensation is determined by the amplitude spectrum 
of the variations of the ambient conditions.
In any case, the study of triggered condensation of warm gases close
to the critical  state can be justified by the self-regulatory hypothesis. 
It is to be noticed that the scale of the inhibitory process is expected to be
much larger than the scale of the triggering processes: a fact that apparently
is a common characteristic of many dynamical systems where spiral structures 
arise (Smolin 1996).  

Triggering mechanisms of star formation are usually related to compression 
and pushing of the ISM gas by high pressure events associated to stars formed 
previously 
(reviews on this topic can be found in Elmegreen 1992, Franco 1992). 
Also, compression of the warm gas by the spiral density wave
have been evocated as a main trigger of its condensation 
(Roberts 1969);
however, stimulated condensation by ionizing flashes may also enter  
as an initiator of the chain of processes that finally results in the 
formation of stars. 
As it will be shown bellow, this stimulating mechanism is particularly 
efficient when the warm gas state is close to its marginal state for 
spontaneous phase transition.
Moreover, variations of the ionizing flux are expected to precede pressure
variations if both variations are associated to the same perturbing event.
Large local variation of the ionizing rate are expected to be present
in the interstellar medium.
Sudden appearance and disappearance of ionizing sources occur 
continuously in the galactic plane.
Also, ``rapid'' variation of the opacity to the ionizing radiation  in
a line of seeing is expected.
Finally, variations of the cosmic ray flux are expected to be present
due to variations of the sources (cosmic ray acceleration in shock fronts
with oblique B-fields; Blandford and Ostriker 1978), and to variations of the 
magnetic field topology (i.e. focalization or dis-focalization of the cosmic
ray stream in a region). It was also proposed 
(Ko \& Parker 1989; Nozakura 1993) 
that star formation controls dynamo activities and hence large scale magnetic 
fields of disk galaxies.

The cooling of the diffuse warm interstellar gas is mainly due to electron
collisional excitation of a) fine structure levels and metastable states of
the positive ions $C^+$, $Si^+$, $Fe^+$, and $S^+$, and b) $Ly_{\alpha}$
excitation.
Therefore, an increase of the ionizing flux (and the consequent increase
of the electron density) tends to reduce the gas
temperature if the increase of the cooling efficiency overcomes the
associated increase of the heating. In general, this is the case for
low equilibrium ionization fractions when the kinetic energy of the
electrons that result from the ionization process is small.
Depending on the value of the pressure and on the increase of
the ionizing flux, the new thermal equilibrium state might be located in
the cool neutral branch. That is: sufficiently large variations of the
ionizing rate are expected to provoke the warm to cool phase transition.
Moreover, if the ionizing flux variation has a flash-like variation, then,
after the passing of the flash, the cooling rate remains enhanced due to
the inertia of ionization. 

In order to illustrate this effect, we consider variations of the
primary ionization rate by cosmic rays $\zeta_o$ which is the
accepted main source of ionization of the interstellar warm gas far away
from massive stars. Other mechanisms of ionization have been proposed
(i.e. OB stars (Reinolds \& Cox 1992), 
the neutrino decay theory (Sciama 1990, 1993)), 
but for the present analysis
the exact mechanism of ionization is not relevant. What is important here
is the variation of the ionizing and heating rates due to a change of
the considered ionizing flux.

\setcounter{figure}{0}
\renewcommand{\thefigure}{5}
\begin{figure}[htb]
\vspace{7.8cm}
\includegraphics{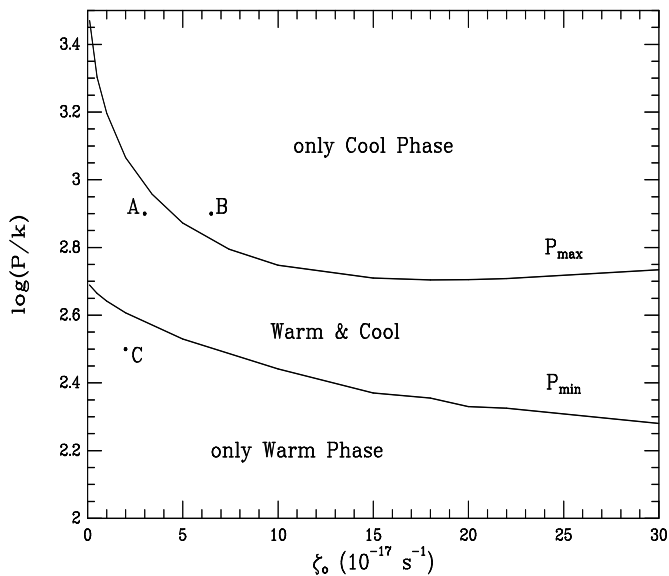}
\caption[]{The maximal ($P_{max}$) and the minimal ($P_{min}$) pressures of the
thermo-chemical equilibrium curve plotted as a function of $\zeta_o$ (see text).
}
\end{figure}

The main changes of the curve of thermo-chemical
equilibrium (TCE) when $\zeta_o$ is changed are summarized in Fig. (5), where
the maximal ($P_{max}$) and the minimal ($P_{min}$) pressures of the TCE 
curve are plotted as a function of $\zeta_o$.
All the results in this section correspond to the standard solar neighborhood
far UV energy density and gas composition.
The labels are used to remind that if $P<P_{min}$
(or $P>P_{max}$) then there is only one possible state of TCE
in the diffuse and warm phase (or in the dense and cool phase).
If $P_{min}<P<P_{max}$ then the gas may reach any of the two stable branches
(the warm or the cool branch).
If the gas is initially in TCE at the warm branch (for example with the
external conditions ($P_o,\zeta_{oA}$) corresponding to the point (A) in
Fig. (5), then, the gas could be forced to evolve toward the cool branch
if the external conditions are changed, for example, to ($P_o,\zeta_{oB}$)
corresponding to the point (B).
The typical time for the transition from the warm to the cool branch is
$\sim 10^7$ yr.
Once the gas reaches the cool phase, the external conditions can change
again to ($P_o,\zeta_{oA}$) but the gas will remain in the cool branch.
To drive the gas to the warm phase again, the external conditions
must be changed, for example, to ($P_{oC},\zeta_{oC}$)  corresponding to
the point (C) in Fig. (5).
Now, the gas could return to the initial state if the external
conditions changes to the initial condition ($P_o,\zeta_{oA}$).

The phase transitions described above assume that the time
between the consecutive changes of $\zeta_o$ are long enough to reach
TCE.
If the variation of $\zeta_o$ occurs before TCE is achieved, then, the
phase transition does not necessarily occur.
Here we will consider the effect of flash-like variations of $\zeta_o$ at
constant pressure.
That is, at the beginning we assume that the gas is in the warm branch
in the equilibrium state corresponding to the external conditions
($P_o,\zeta_{oa}$).
Then, $\zeta_{o}$ is increased abruptly by a factor $F_z$ during a lapse
of time $\Delta_t$, after which the primary ionization rate by cosmic rays
returns to the initial value $\zeta_{oa}$.
As mentioned above, a fact
that favors the phase transition is that the recombination
time is much larger than the cooling time; then after the end of
an ionizing pulse, the cooling rate remains enhanced due to the 
non-equilibrium excess of electrons.

In this simple analysis we
neglect non-local processes as thermal conduction, cosmic ray attenuation,
radiative transfer, and gas dynamics. Only local processes of radiative
cooling, heating, H ionization and recombination are
considered (Parravano 1986), assuming that the pressure remains constant 
during evolution.
The neglected non-local processes may play an important role in the local
evolution of the gas and in should determine the spatial variations
of the physical variables. However, here we are interestested in showing
that the inertia of ionization (after the passing of the ionizing pulse) can,
in many cases, enhance sufficiently the cooling efficiency to produce a phase 
transition from the warm gas phase to the cool phase.
The evolution of the temperature and the ionization degree is calculated by
solving simultaneously the energy conservation equation (2), and the kinetic 
equation (1) restricted to the Hydrogen ionization-recombination processes.

\setcounter{figure}{0}
\renewcommand{\thefigure}{6}
\begin{figure}[ht]
\vspace{7.6cm}
\includegraphics{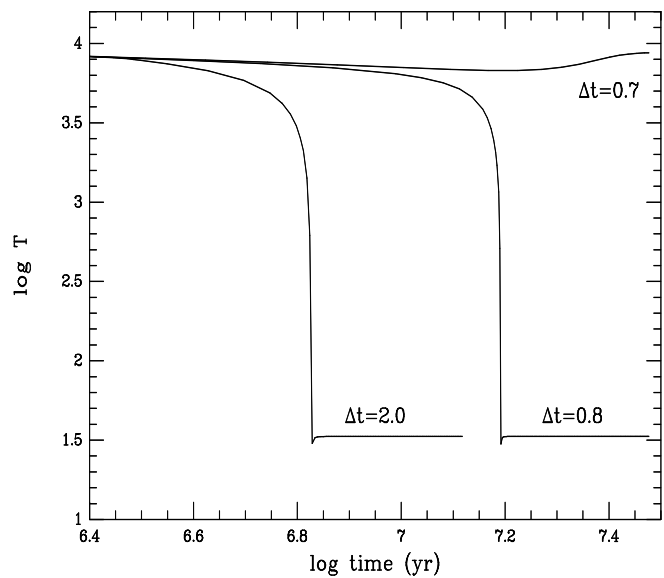}
\caption[]{The temperature evolution
of the warm gas initially in thermo-chemical equilibrium
close to the critical state for spontaneous condensation.
The three curves correspond to the evolution for three labeled
values of the pulse duration when $F_z=20$.
}
\end{figure}
\setcounter{figure}{0}
\renewcommand{\thefigure}{7}
\begin{figure}[htb]
\vspace{7.6cm}
\includegraphics{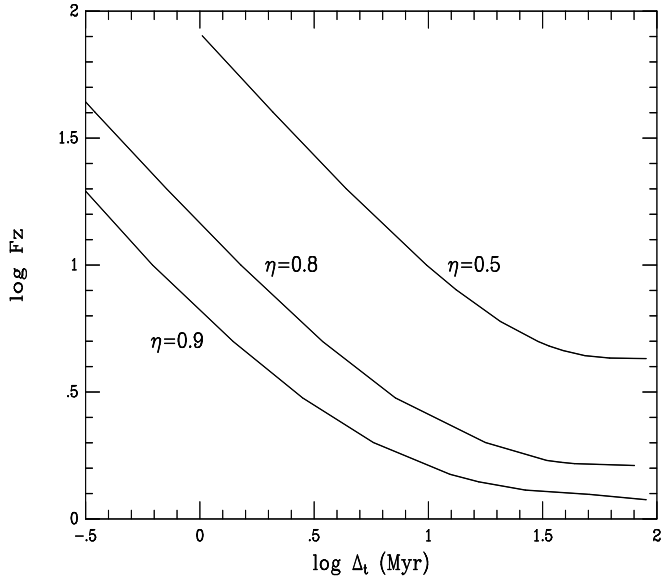}
\caption[]{The critical values of $F_z$ as function of $\Delta_t$
for the three labeled values of $\eta$.
}
\end{figure}

In order to show the effect of ionizing pulses, Fig. (6) shows the evolution 
at constant pressure of a warm gas initially in thermo-chemical equilibrium 
close to the critical state for spontaneous condensation. The pressure $P_{o}$
is bellow $P_{max}$ by a factor $\eta=P_{o}/P_{max}(\zeta_{oo})=0.8$, when
the pre-flash primary ionization rate is $\zeta_{oo}=10^{-17} \mbox{\rm s}^{-1}$.
The three curves in Fig. (6) correspond to the evolution for three different 
values of the pulse duration $\Delta_t=0.7$, $0.8$, and $2.0$ Myr, when the 
flash is initiated at $t=10^6$ yr and the factor of increment of $\zeta_{o}$ 
is $F_z=20$.  
Note in Fig. (6) that there is a critical value of $\Delta_t$ bellow which 
condensation does not occur (in this case $\sim 0.7$ Myr). 
Note also that the time required to complete the phase transition decreases 
as $\Delta_t$ increase, but the time required for the temperature drop from 
$\sim 5000 \mbox{\rm K}$ to $\sim 50 \mbox{\rm K}$ is insensitive to $\Delta_t$. 
The kink at the bottom of the $\Delta_t= 0.8$ and $2.0$ curves, and  the smooth 
dip in the $\Delta_t=0.7$ curve are due to the inertia of the ionization 
fraction (the recombination time is much longer for the conditions 
corresponding to the upper curve).

In Fig. (7) the critical values of $F_z$ are plotted as function of $\Delta_t$
for the three labeled values of $\eta$. These curves divide the plane 
$(F_z,\Delta_t)$ in two regions: above the curves the flash is capable 
of inducing phase transition, and bellow the curve the gas returns to 
its initial state after the passing of the flash. 
Note that a cosmic ray pulse with a factor of increment $F_z$ of 
the order of $10$, and a duration of about one Myr can induce 
the condensation of the warm gas in TCE at $\eta > 0.8$.

\section{Summary and Conclusions}

The variation of the ionizing flux as a mechanism for stimulating the
condensation of the diffuse gas was considered. To illustrate this effect,
two situations were examined: one on the context of pregalactic conditions
(a free metal cloud), and the other on the context of the actual
interstellar medium (a gas with solar abundances).
We have focused our attention on flash-like variations; that is,
during a ``short'' period of time the ionizing flux is enhanced in
comparison to the pre and post flash values.
In both cases the cause of the induced phase change is the same:
the enhancement of the cooling rate by the increase of the electron density
caused by the momentary increase of ionizing flux.
After the passing of the flash, the cooling rate remains
enhanced due to the inertia of the ionization.
In the first case (metal free gas) 
we show that after the lapse of UV flux decrement, recombination
continues, but at a lower rate than the cooling.
The excess of electrons at these relatively low temperatures results
in an enhancement of the $H_2$ rate formation due to the enhanced
abundance of the $H^-$ intermediary.
Even when $\epsilon _{2} > \epsilon _{0}$, the abundance of $H_2$ may reach
a large enough value to produce considerable self-shielding.
If the cloud reaches a critical value ($\tau \sim 1$) for the optical depth
at dissociating
frequencies, the $H_2$ abundance grows very fast, allowing the cooling
of the cloud to temperatures of the order of $100 \mbox{\rm K}$.
The temperature drop occurs in a fraction ($\sim 0.2$) of the free fall time
provoking a rapid decrease of the Jeans mass.
However, if the post-flash dissociating level is large enough the $H_2$
formation can be inhibit and the cloud remains warm. 

In the second case (solar abundances gas) the dominant cooling mechanism
of the warm neutral gas (the excitation of heavy ions by
electron impacts) is proportional to the electron density,
and therefore, the ionizing flash increases the cooling efficiency. 
We considers flash-like variations of the primary ionization rate
by cosmic rays and calculate the marginal flash characteristics to induce
warm gas condensation. 
We show that, for the expected states of the warm interstellar gas, 
ionizing flashes may easily induce the phase transition from the warm to the cool
phase.
The phase transition is completed 
in about $10^7$ yr; however, the drop in the temperature from $\sim 5000 \mbox{\rm K}$
to $\sim 50 \mbox{\rm K}$ occurs in about $5\times 10^5$ yr.

The results indicate that the mechanism of induced condensation studied here
might play a relevant role in the gas evolution of the diffuse gas in both, 
the pregalactic and the actual interstellar medium conditions.

The above results include only local processes. However, non-local processes
like thermal conduction impose restrictions on the size of the condensing 
structures. Thermal conduction tends to attenuate temperature gradients, and
therefore, it imposes a minimal value for the mass of the condensing 
structures 
(Corbelli \& Ferrara 1995;
Ib\'{a}\~{n}ez \& Rosenzweig 1995; 
Steele \& Ib\'{a}\~{n}ez 1997). 
Critical masses of the order of galaxy masses are obtained in the case of a metal-free gas
(Ib\'{a}\~{n}ez \& Parravano 1983) and in the case of an actual interstellar gas, masses
of the order of one solar mass are obtained 
(Parravano 1986;
Parravano, Ib\'{a}\~{n}ez \& Mendoza 1993). 
However, a wide range of critical masses are obtained depending on the initial gas state, 
and ambient conditions.
Therefore, in addition to the restrictions on $F_z$ and $\Delta_t$
to induce the warm gas condensation, the cosmic ray flash should cover
a region greater than the critical value imposed by thermal conduction
and other diffusive processes. 

\begin{acknowledgements}
We are very greatful to the referee A. Ferrara for his useful comments that
motivated us to include Helium in the metal free model and to clarify
the discussion by including the phase diagram.
This work has been supported by CDCHT and calculations were performed at CECALCULA,
both institutuons of the Universidad de Los Andes.
\end{acknowledgements}

\end{document}